\documentclass{PoS}
\usepackage{booktabs}

\makeatletter
\let\MYcaption\@makecaption
\makeatother
\usepackage{subcaption}
\makeatletter
\let\@makecaption\MYcaption
\makeatother

\title{Electronics Development for \\the New Photo-Detectors (PDOM and D-Egg) \\ for IceCube-Upgrade}

\ShortTitle{PDOM and D-Egg Electronics for IceCube-Upgrade}

\author{
The IceCube Collaboration\footnote{For collaboration list, see PoS(ICRC2019) 1177.}\\
{\itshape \href{http://icecube.wisc.edu/collaboration/authors/icrc19_icecube}{http://icecube.wisc.edu/collaboration/authors/icrc19\_icecube}}\\
E-mail: \email{rnagai@hepburn.s.chiba-u.ac.jp}
}

\abstract{The planned IceCube-Upgrade will enhance the capability of IceCube in the detection of GeV-scale neutrino physics and enable an improved measurement of the properties of the glacial ice. 
Three types of new optical sensors will be deployed during the Upgrade: PDOM, D-Egg, and mDOM. 
Since the design of the PDOM and D-Egg are very similar, the development of the front-end electronics for the two optical sensors has been merged. 
The photo-electron signals detected by the PMTs are digitized with high-speed ultra-low power ADCs and processed in an FPGA, before being sent to the data acquisition system located on the surface of the South-Pole glacier. 
The almost final revision of the front-end electronics is equipped with the common microcontroller unit and the communication daughter board for simplifying the communication scheme for the three different modules. 

This contribution focuses on the design of the front-end electronics and presents first results from the performance tests.

\vspace{4mm}
{\bfseries Corresponding author:}
Ryo Nagai$^{1}$, \speaker{Aya Ishihara}\,$^{1}$, and the IceCube Sensor Electronics Group, \\
{$^{1}$ \itshape Department of Physics and Institute for Global Prominent Research, Chiba University, Chiba 263-8522, Japan}\\

}

\FullConference{36th International Cosmic Ray Conference -ICRC2019-\\
		July 24th - August 1st, 2019\\
		Madison, WI, U.S.A.}

\begin{document}
\section{Introduction} 
%IceCube Neutrino Observatory~\cite{icecube}, a cubic-kilometer neutrino detector in Antarctic glacier, is the largest neutrino detector in the world. 
%IceCube-Gen2~\cite{ic2}, the next generation neutrino observatory, is the extension of IceCube. 
The next generation neutrino observatory named IceCube-Gen2~\cite{ic2} is the extension project of IceCube.
IceCube-Gen2 is expected to expand its neutrino detection sensitivity for GeV--PeV energy neutrinos significantly. 
It will occupy $\sim$8~km$^3$ of Antarctic ice, surrounding the current IceCube instrumentation area, 
and consist of an array of $\sim$10,000 digital optical modules (DOMs) to capture Cherenkov photons efficiently from the secondary charged particles produced in the neutrino interactions. 
Prior to the main high energy array extension, the IceCube-Upgrade array will be installed during the 2022/2023 South-Pole season. 
%The upcoming project, IceCube-Upgrade, will be constructed during the 2022/2023 South-Pole season. 
It
%The IceCube-Upgrade array 
will consist of seven densely instrumented strings in the center of the current IceCube array. 
The Upgrade will significantly enhance IceCube's GeV-scale neutrino physics capabilities.  
In addition, the IceCube-Upgrade will host new calibration devices to measure the optical properties of glacial ice with higher precision, thereby enabling an improved reconstruction of neutrino-induced events. 
%The 

Three types of new optical sensors will be deployed within the IceCube-Upgrade; mDOM~\cite{mDOM}, D-Egg~\cite{DEgg}, and PDOM~\cite{PDOM}. 
Each string will hold $\sim$100 devices, with a mixture of the three types of the optical sensors. 
The powering quad cables connect 
the optical sensors installed in the deep ice to the data acquisition (DAQ) system located on the surface of the South-Pole glacier. 
They provide $\pm$48~V power, along with the operation commands with $\sim$2~MHz. 
The commands and collected data are digitized to avoid data loss over the $\sim$2,500~m powering cable. 
Each module generates the 5~V power from the external $\pm 48$~V power. 
The power consumption for each module is strictly limited to less than 4~W. 
In addition, the modules must work stably at the temperature range of $-40^{\circ}$C to $-20^{\circ}$C. 
To take advantage of the photomultiplier tube (PMT) range from less than 1~p{.}e{.} to the saturation region, 
the dynamic range of the digitization have to be kept wider, e{.}g{.} 0--4~V. 
The electrical noise should be suppressed within $\sim$0.5~mV. 

Although each photo-sensor has been developed separately from different groups within the IceCube Collaboration, these three R\&D sensors share the same development principles. 
The analog signals from the PMTs are digitized at the mainboard. 
The processed signals are sent to the same surface DAQ system, and thus
their electronics are required to receive and transmit data with the same format, 
simplifying the communication with the surface DAQ system. 
As this contribution focuses on the PDOM and D-Egg, we summarize the specifications for them in the following paragraphs. 

\begin{figure}[t]
\centering
\begin{minipage}{.6\textwidth}
\centering
\includegraphics[height=.19\textheight]{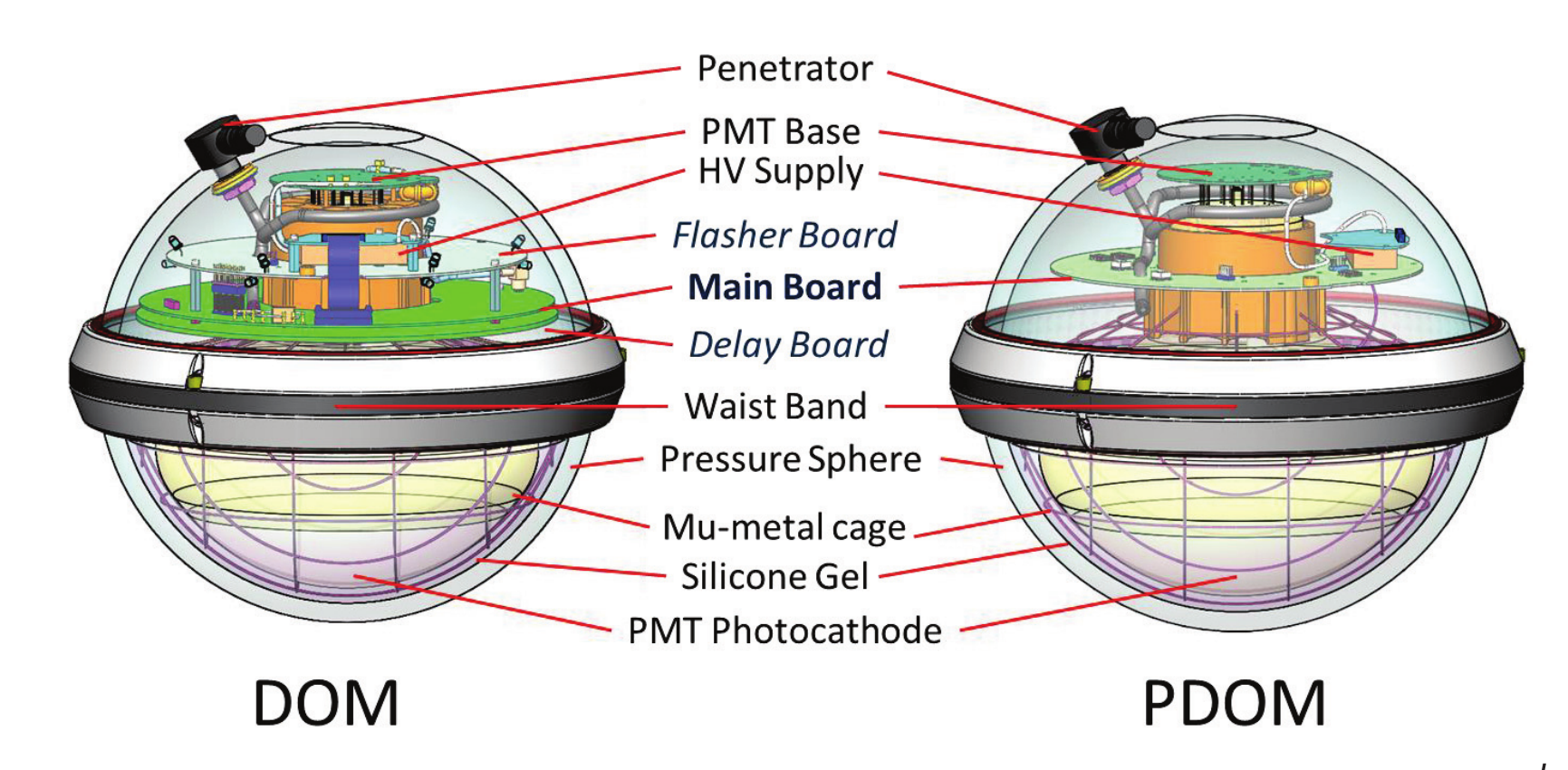}
\subcaption{IceCube DOM and PDOM~\cite{PDOM}}\label{fig:pdom}
\end{minipage}~
\begin{minipage}{.38\textwidth}
\centering
\includegraphics[height=.19\textheight]{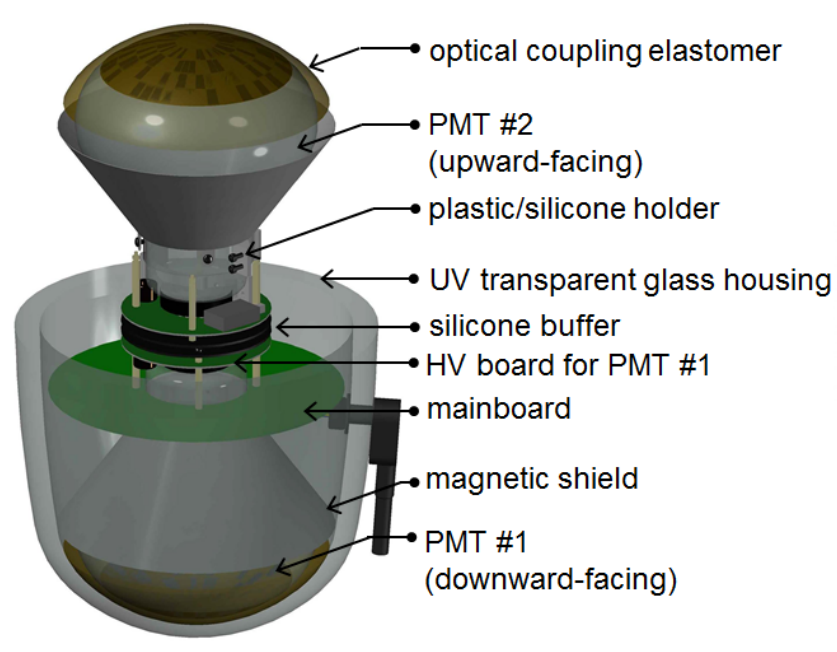}
\subcaption{D-Egg~\cite{glass}}\label{fig:degg}
\end{minipage}
\caption{Views of the IceCube DOM~(a), PDOM~(a), and D-Egg~(b). IceCube DOM and PDOM utilize the same pressure vessel and the internal structures except for the electronics. D-Egg houses two 8-inch PMTs along with a mainboard. }
\end{figure}

%\subsection{PDOM}
The PDOM concept is based on the current IceCube DOM incorporating upgraded electronics (Fig.~\ref{fig:pdom}).  
It houses one down-facing 10-inch high quantum efficient PMT, readout electronics, and calibration devices inside 12-inch glass sphere. 
A high-performance and low-power ADC allows for continuous digitization of the PMT signals, thereby eliminating data-taking dead-time and enabling flexible data-taking, while the waveform window of the current IceCube DOM is limited to at most 6.4~$\mu$s. 
%he analog delay board will be removed to shorten the length of the analog signal line. 

%\subsection{D-Egg}
The ``D-Egg'' (Dual channel optical sensors with Ellipsoid Glass for IceCube-Gen2) design features two 8-inch high quantum efficient PMTs facing up and down, respectively (Fig.\ref{fig:degg}). 
The PMTs are housed inside an ellipsoidal pressure vessel made of UV transparent borosilicate glass. 
The shape of the glass vessel has been optimized for both transparency and to withstand high pressure during freeze-in~\cite{glass}. 
%While the PMTs in a D-Egg are slightly smaller than the one in the PDOM, the photosensitive areas are comparable. 
While the PMTs in a D-Egg are slightly smaller than the one in the current IceCube DOM, 
the photon detection performance of D-Egg will be twice that of the current IceCube DOM~\cite{deggperf}. 
Furthermore, $4\pi$ sensitivity is achieved through the up-facing PMT enabling collection of photons from any direction. 
The data-taking scheme is same as the one of the PDOM~\cite{PD18}. 
The waveform from the two PMTs is digitized at two ADCs and are sent to an FPGA. 
Data processing is performed continuously. 

\section{Design of the Front-End Electronics} 
The main features of the front-end circuit are handling the waveform data from PMTs and sending it to the surface DAQ system. 
The mainboard receives the power as well as the commands from the quad cables. 
%The received commands are sent to the processor or the FPGA, 
%then the operation application installed in the processor or the FPGA is operated. 
Collected data is digitized at the ADC located on the most front-end part of the mainboard. 
The hardware triggers are published based on the digitized data at the FPGA. 
The triggered data will be sent to the processor, 
which converts them to the data flow and sends them to the surface DAQ system. 
Software triggers are added at this stage. 

The final design of the electronics is based on the first prototypes for the PDOM and D-Egg. 
The first PDOM prototype mainboard was produced in 2016, 
and that of the D-Egg was produced in 2017. 
Both have been developed separately. 
The functionality is identical, while the schematics are different from each other. 
We tested their performances with respect to the stability of the waveform reconstruction, 
the noise level, and the operation stability at low temperature. 
We confirmed that they satisfied the requirements except for the noise level. 
The results of the performance tests for the D-Egg first prototype mainboard is summarized in~\cite{PD18}. 

\begin{figure}[t]
\centering
\includegraphics[width=.95\textwidth]{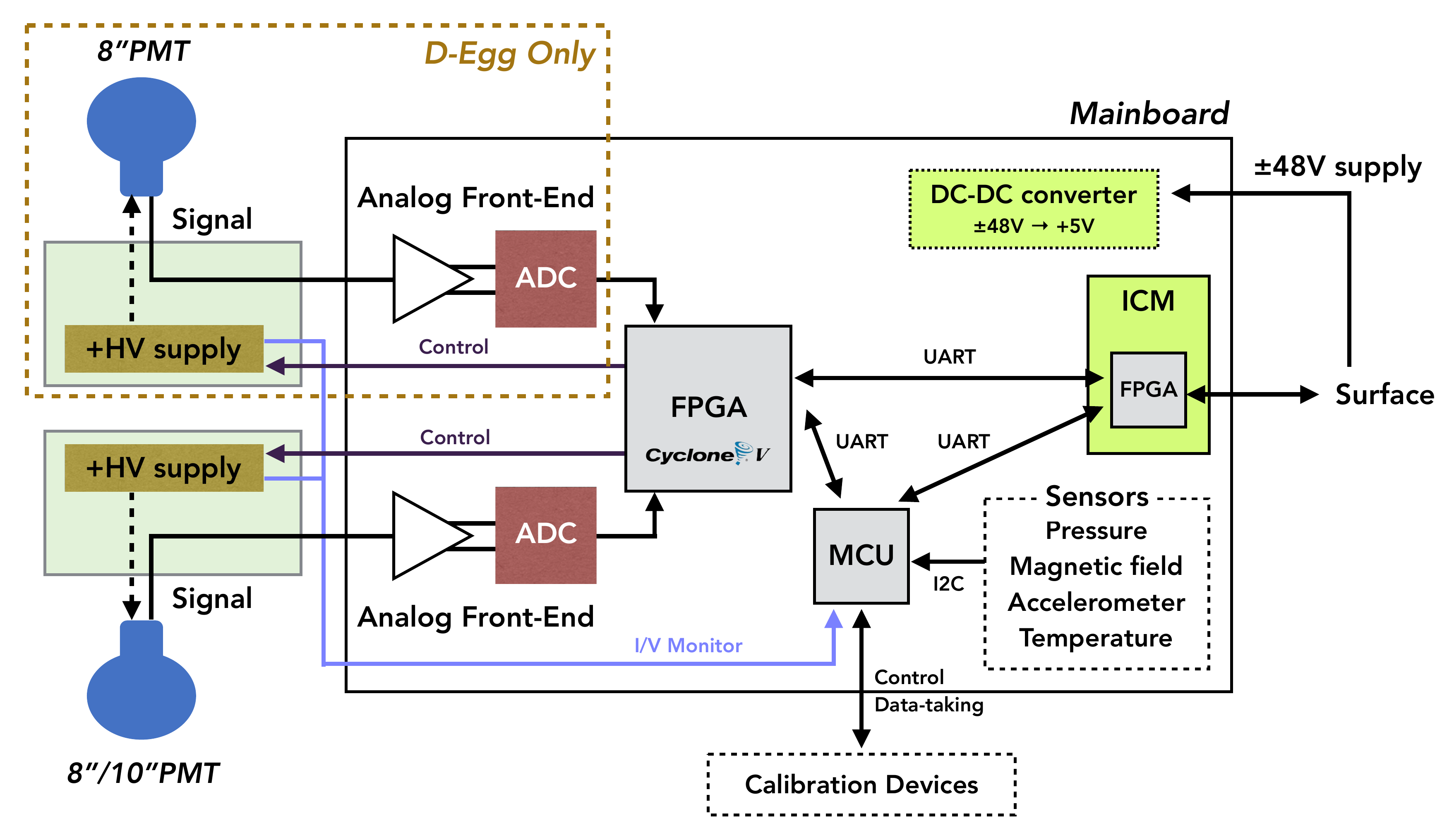}
\caption{Simple block diagram for the PDOM/D-Egg mainboard. The upper PMT system is only for the D-Egg. 
The mainboard contains an FPGA, a microcontroller unit (MCU) as the processor, and the separated board ``IceCube Communication Module'' (ICM). 
ICM receives commands from the surface, generates commands to the FPGA and the MCU, and sends collected data to the surface DAQ system. 
The MCU drives the operation software. The FPGA receives digitized data from the ADC and sends them to ICM. 
The board power is supplied by the surface system, and the voltage is stepped down to 5~V to supply voltage to the mainboard components.  } \label{fig:blockdiag}
\end{figure}

\begin{figure}[t]
\centering
\includegraphics[width=.75\textwidth]{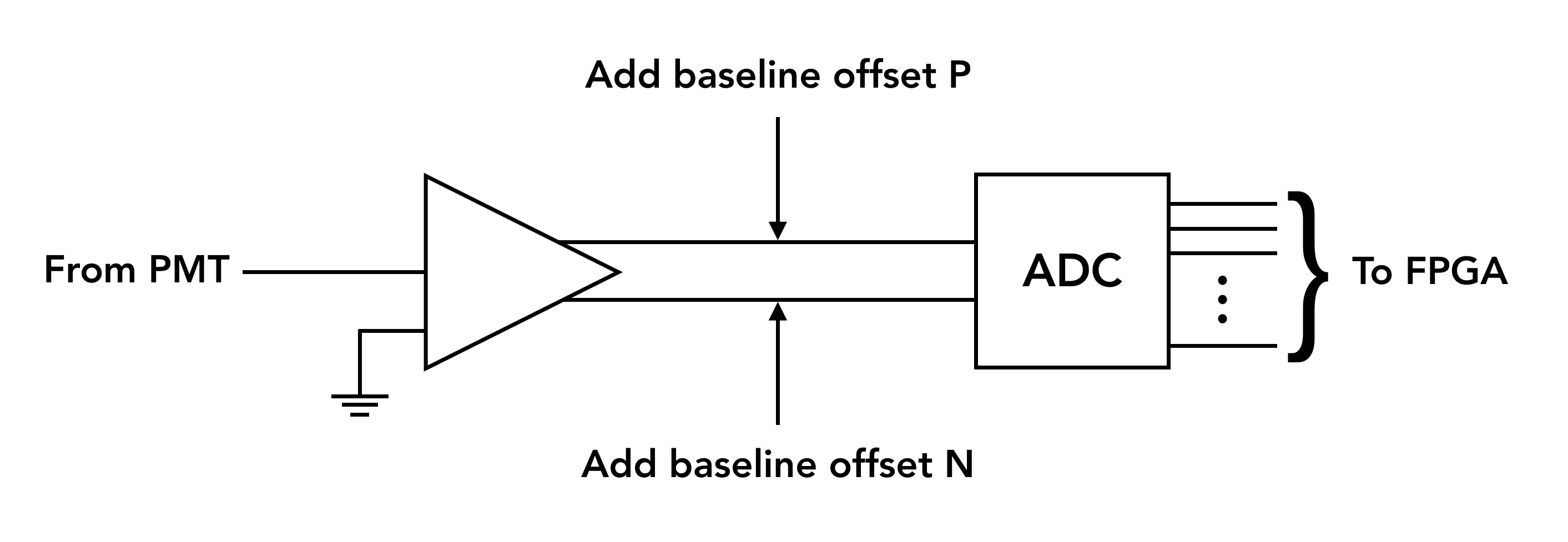}
\caption{Analog front-end circuit. An amplifier generates a differential signal. 
The proper baseline offset is added before the ADC to keep the information of the undershoot of the waveform. The offset value is controlled by the FPGA. 
} \label{fig:afe}
\end{figure}

Since the data-taking concepts of the PDOM and D-Egg are similar, 
the schematics of the PDOM and D-Egg mainboards are unified. 
It contributes to reduce human resources needed to develop and maintain an on-board firmware/software. 
The main functions for the PDOM/D-Egg mainboard are the following: 
\vspace{-.4\baselineskip}
\begin{itemize}
\setlength{\parskip}{2pt}
\setlength{\itemsep}{0pt}
\item Analog pulse shaping 
\item Digitization 
\item Slow Control
\item Operating data-taking and detector calibration 
\item Communication with the surface DAQ system
\end{itemize}
\vspace{-.4\baselineskip}
A simplified block diagram for the PDOM/D-Egg mainboard is shown in Fig.~\ref{fig:blockdiag}.
The waveform taken by the PMTs is converted to a differential signal at the amplifier on the analog front-end circuit (Fig.~\ref{fig:afe}). 
The amplification rate is optimized for a signal voltage range of 0.5~mV to 3.5~V, which corresponds to 
0.1~p{.}e{.} to $\sim$200~p{.}e{.} at $10^7$ PMT gain, to detect both the single photo-electron signal with $\sim$30 ADC counts and the PMT saturation signals. 

There is a toroidal coil between the PMT and the signal line to the mainboard to cut off the PMT high voltage to the signal line, because the positive high voltage is applied to the PMT. 
The toroidal coil can induce undershoot, especially when high amplitude pulse is observed. 
An additional baseline offset is added to the differential waveform to keep the information of the undershoot.   
The standard value of the baseline offset is $\sim$2000 ADC counts, which corresponds to $\sim$500~mV. 
The offset value should be optimized for the level of the undershoot. 
The study of the undershoot is ongoing. 
We can control the baseline offset from the FPGA via the SPI-controllable 16-bit DAC. 

The high-performance low-power ADC receives the differential signal with an additional baseline offset. 
Data is continuously digitized by the 14-bit ADC at 250~MSPS. 
This is enough time resolution ($\sim$~few~ns) to perform the track reconstruction in IceCube. 
Digitized data is then processed on the FPGA (ALTERA Cyclone-V E 5 series) on the mainboard.   
It is slightly smaller and cheaper than the current IceCube readout FPGA (ALTERA Excalibur EPXA-4), 
as several main functions have been transferred from the FPGA to the IceCube Communication Module (ICM) and the microcontroller unit (MCU) as a common processor. 
The FPGA stores data, publishes the triggers, and sends the triggered signal to the ICM or the MCU. 
The FPGA also controls the signal delay using the shift registers, which are controlled with the analog delay lines for the IceCube DOM mainboard. 
All electronic components are selected with taking into account the power consumption. 
The total power consumption of the prototype mainboard is estimated to be $\sim$2.5~W, 
though it depends on the FPGA/MCU logic. 

%\subsection{MCU and ICM}
The new components, the MCU and the ICM, take over the functions from the FPGA. 
They are developed as common devices for the mDOM, D-Egg and PDOM to enable a common application. 
We describe the detail of those new components in the following subsections. 
\subsection{MCU}
The high-performance ARM microcontroller, STM32H743, is adopted as an on-board MCU. 
It operates the common software (written in C),  
which is responsible for the configuration of the FPGA, processing triggered data from the FPGA, 
driving DOM calibration devices, calibrating PMTs and ADCs, 
and monitoring the high voltage supply for PMTs and various I2C-readable sensors on the mainboard (pressure, temperature, accelerometer, and magnetic field). 
These on-board sensors measure the environment inside the vessel, which is sent to the surface for the offline calibration. 
It has been developed based on the current IceCube software~\cite{sw}, which has been running 
on the CPU embedded in the FPGA on the IceCube DOM mainboard. 
% (software description)

External calibration devices such as the LED flasher circuit and the camera system are controled by the MCU. 
The mainboard has a 20-pin connector for the driver circuit for the calibration devices. 
These dedicated devices are connected in a daisy chain to reduce the parallel controlling lines. 
The commands and addresses are sent as the SPI signals. 

The calibration software runs in the MCU. 
The PMT gain is calculated using the collected single photo-electron events. 
The calibration data will be sent to the surface through the ICM.

\subsection{ICM}
The ICM is designed as a $35\times 65$~mm$^2$ daughter board to be attached to the mainboard. 
It has 8 layers and houses a Xilinx FPGA Spartan-7, and is responsible for communication with the surface DAQ system and time calibration for the front-end module. 
The ICM receives commands from the surface DAQ system and converts the commands with the Universal Asynchronous Receiver/Transmitter (UART) format at the Xilinx FPGA. 
The interlock signals, which enable the FPGA, the MCU, the high voltage supply, and the calibration devices, 
are published. 
These components can be turned off from the surface system in an emergency through the interlock signal. 
The first prototype of the ICM was completed, though the firmware for the ICM is under development.

\section{Prototyping for D-Egg}
\begin{figure}[t]
\centering
\includegraphics[width=.45\textwidth]{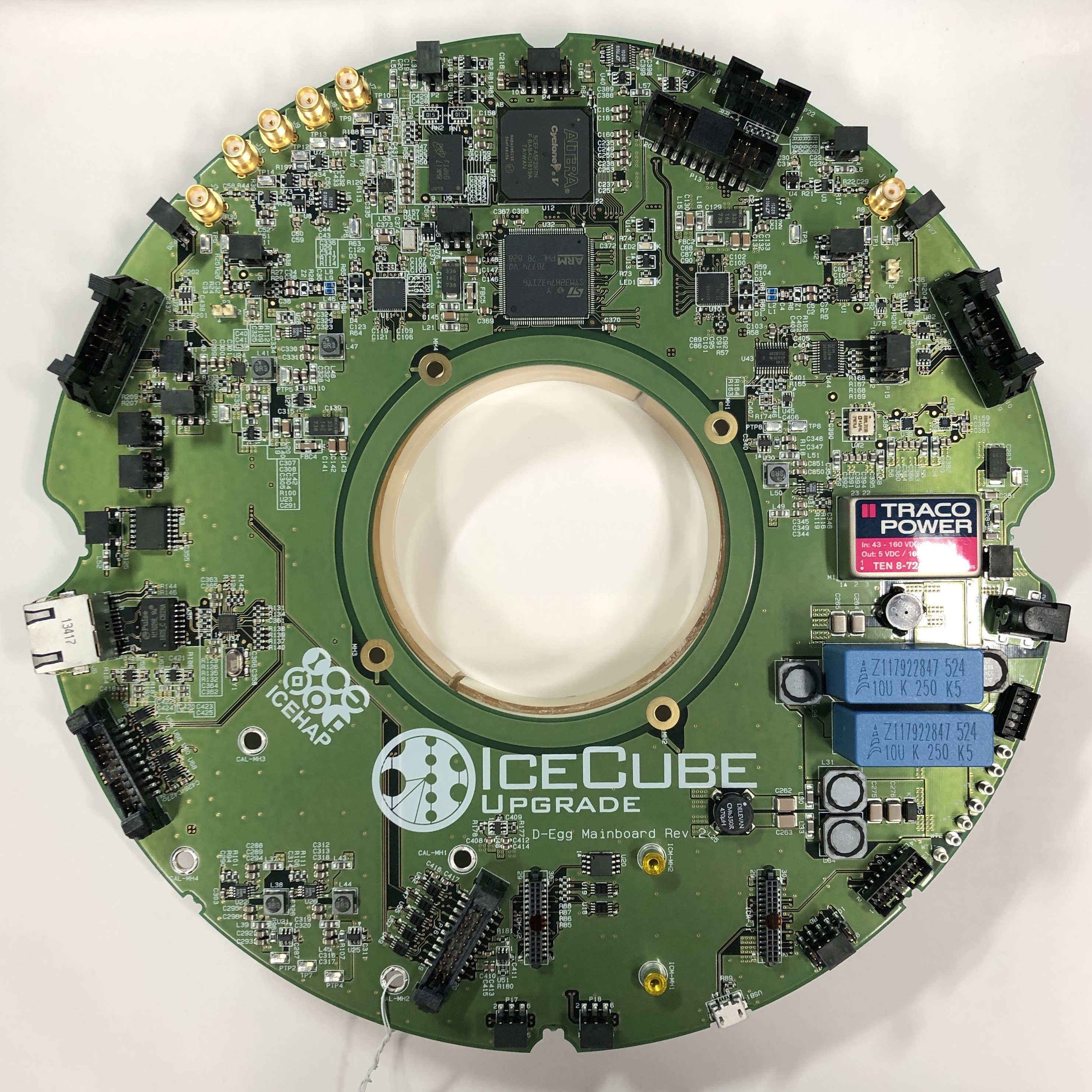}~
\includegraphics[width=.45\textwidth]{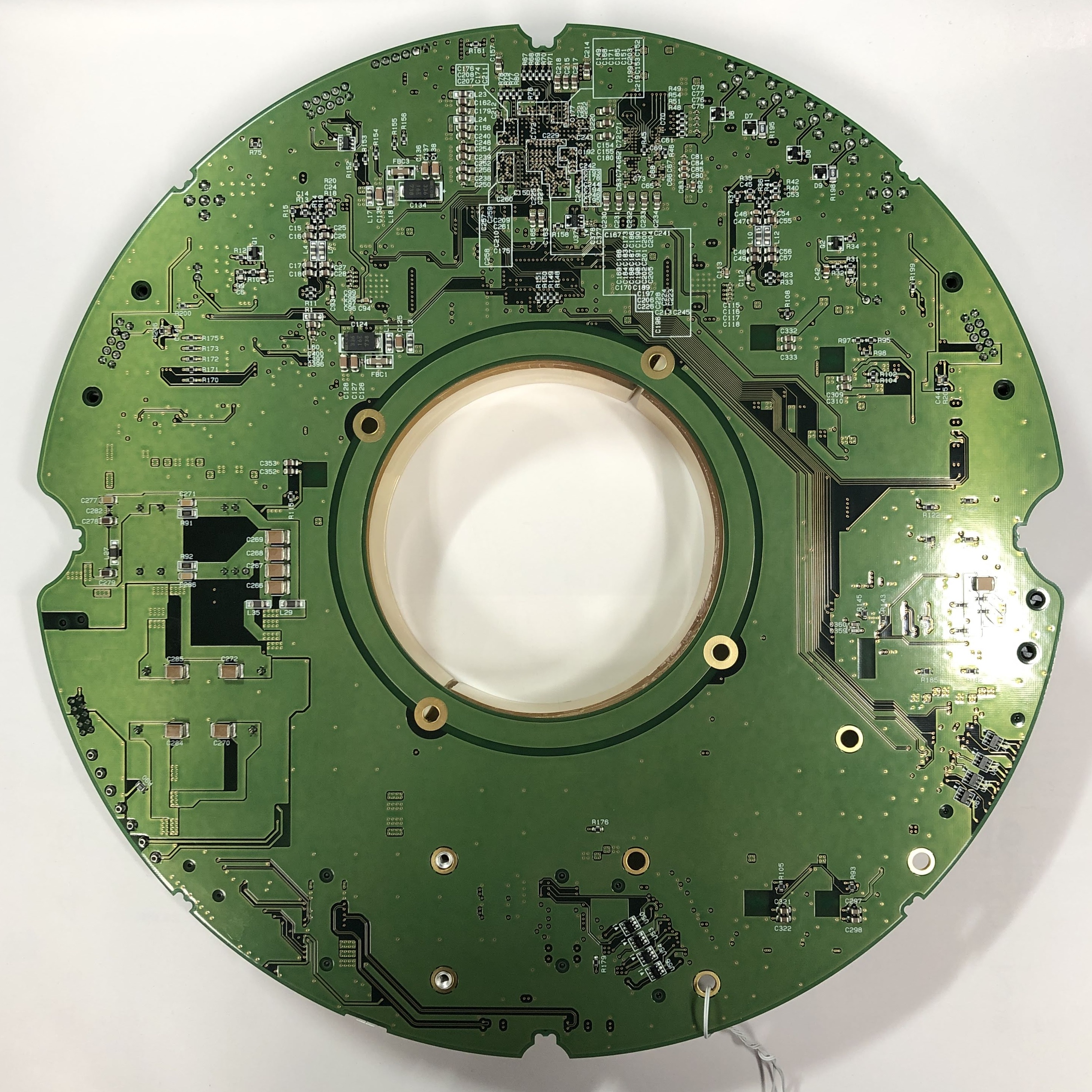}
\caption{Photos of the top side (left) and bottom side (right) of the D-Egg mainboard prototype.}\label{fig:DEggRev2}
\end{figure}

While the schematics for the PDOM/D-Egg mainboards are entirely identical, 
the board layout and routing have been done separately because of different mechanical requirements. 
The D-Egg mainboard prototype was produced in June, 2019 (Fig.~\ref{fig:DEggRev2}). 
The layout and routing have been completed, accounting for the D-Egg spacial constraints and suppression of noise. 
The prototype board has 12 layers of 1.6~mm thickness FR-4. 
The radii of the outer/inner edges are 246~mm and 86~mm, respectively. 

Signal or GND/power plane layers are assigned to take into account the noise, 
because a high noise level ($\sim$1.2~mV against the single photo-electron signal of $\sim$8~mV~\cite{PD18}) was observed in the first prototype of the D-Egg mainboard. 
The GND/power layers are sandwiched between two signal layers to suppress noise from the return current. 
Also, 
noise from the system clock feed lines has been observed for the first prototype circuit board.
%Fig.~\ref{fig:fft} shows the FFT plot of the baseline waveform taken by the first prototype. 
The multiple 20~MHz frequent noise was observed in the baseline waveform taken by the first prototype. 
It corresponds to the oscillator frequency of the system clock for the first prototype. 
As we looked at the routing of the first prototype board, the clock feed line came across the analog front-end circuit. 
For the final revision, the clock lines are routed to avoid the region of the analog front-end circuit. 

Connectors are located only on the top side of the board for the assembling reasons. 
Some cables come from the back side of the board, thus we prepare 
cut-outs to put them to the top side. 
Powering has been tested using two AC-DC converters (Mean Well MSP-200-48), which supply $\pm$48\,VDC. 
Eight test points were prepared. 
Each point is placed near the corresponding DC-DC converter on the board. 
Tab.~\ref{tab:DCDC} shows the result of the powering test. 
No critical problems were observed in this test. 
\begin{table}[tb]
\centering
\caption{Result of the powering test for 4 D-Egg prototype mainboards. No critical problems were observed. }
\label{tab:DCDC} \vspace{1ex}
\begin{tabular}{c|c|cccc}
\toprule
Test Point \# & Designed Value & No.1 & No.2 & No.3 & No.4 \\
\midrule
1 & 5.00~V & 4.96~V & 4.97~V & 4.97~V & 5.00~V \\
2 & 2.50~V & 2.49~V & 2.49~V & 2.49~V & 2.49~V \\
3 & 1.10~V & 1.08~V & 1.09~V & 1.09~V & 1.09~V \\
4 & 3.30~V & 3.31~V & 3.32~V & 3.30~V & 3.32~V \\
5 & 1.35~V & 1.34~V & 1.34~V & 1.35~V & 1.34~V \\
6 & 0.675~V & 0.67~V & 0.67~V & 0.68~V & 0.67~V \\
7 & 0.675~V & 0.67~V & 0.67~V & 0.67~V & 0.67~V \\
8 & 1.80~V & 1.81~V & 1.80~V & 1.80~V & 1.80~V \\
\bottomrule
\end{tabular}
\end{table}

The first FPGA/MCU performance test using a simple test firmware is still ongoing. 
We have confirmed that the ADC of each board responds correctly to the baseline set commands by monitoring the SPI output signal of the ADC. 
The following items are going to be tested with the firmware which we are developing: 
\vspace{-.4\baselineskip}
\begin{itemize}
\setlength{\parskip}{2pt}
\setlength{\itemsep}{0pt}
\item Simultaneous data-taking from two input lines
\item Dynamic range 
\item Analog noise level
\item FPGA--MCU--ICM communication
\item Long-term operation at a low temperature ($-40^{\circ}$C)
\item Strength against vibration during the transportation
\item Performances of the various sensors on the board
\end{itemize}
\vspace{-.4\baselineskip}
We will perform these items in a few months. 
The PDOM prototype mainboard is also being developed. 
The layout/routing design is still ongoing. 

\section{Conclusion}
The almost final revision of the PDOM/D-Egg mainboard has been developed. 
A common MCU and ICM are employed 
to unify the data-taking procedure among the three different modules: PDOM, D-Egg, and mDOM.  
For the PDOM and D-Egg, we unified the mainboard schematics because of their similar data-taking requirements. 
The waveform from the PMTs is digitized with a high-performance low-power ADC with 250~MSPS. 
Data is processed by an ALTERA FPGA Cyclone-V and the MCU STM32H743, and is sent to the surface DAQ system via the ICM. %
The MCU controls not only the data-taking scheme, but also the calibration devices and the sensors on the mainboard. 
Common firmware for the MCU has been developed for all three module types. 

The board layout and routing are completed separately for PDOM and D-Egg. 
The prototype of the D-Egg mainboard was completed at the end of June, 2019, 
while that of the PDOM mainboard is under production.  
The basic performance features of the D-Egg prototype have been confirmed.

\end{document}